\documentclass{article}
\usepackage{spconf,amsmath,graphicx}
\usepackage{amsfonts, amssymb}
\usepackage{hyperref}
\usepackage{algorithmic}
\usepackage{algorithm}
\usepackage{xcolor}
\usepackage{multirow}
\usepackage{float}
\usepackage{enumitem}
\setlist{nosep, leftmargin=14pt}

\usepackage{mwe} 

\title{\textbf{CFL-SparseMed:} Communication-Efficient Federated Learning for Medical Imaging with Top-k Sparse Updates}
\name{
	Gousia Habib\raisebox{0.5ex}{\textsuperscript{†*}}\thanks{Equal contribution}\textsuperscript{1,2}, 
	Aniket Bhardwaj\raisebox{0.5ex}{\textsuperscript{†}}\thanks{Equal contribution}\textsuperscript{2}, 
	Ritvik Sharma\textsuperscript{2}, 
	Shoeib Amin Banday\textsuperscript{3},
	Ishfaq Ahmad Malik\textsuperscript{2},
}
\address{
	\textsuperscript{1}University of Helsinki, 
	\textsuperscript{2}Shoolini University,
	\textsuperscript{3}King Khalid University
}
\begin{document}
%
\maketitle
\begin{abstract}
Secure and reliable medical image classification is crucial for effective patient treatment, but centralized models face challenges due to data and privacy concerns. Federated Learning (FL) enables privacy-preserving collaborations but struggles with heterogeneous, non-IID data and high communication costs, especially in large networks. We propose \textbf{CFL-SparseMed}, an FL approach that uses Top-k Sparsification to reduce communication overhead by transmitting only the top k gradients. This unified solution effectively addresses data heterogeneity while maintaining model accuracy. It enhances FL efficiency, preserves privacy, and improves diagnostic accuracy and patient care in non-IID medical imaging settings. The reproducibility source code is available on \href{https://github.com/Aniket2241/APK_contruct}{Github}.
\end{abstract}
\section{Introduction}
\label{sec:intro}

Advanced medical diagnosis increasingly relies on medical imaging analytics. Modern imaging modalities such as MRI, CT, ultrasound, and digital pathology have greatly expanded the volume of medical data \cite{refK15, refK16}. Concurrently, advances in artificial intelligence (AI), especially deep learning, have improved data processing efficiency and clinical decision-making \cite{refK11}. However, applying AI in medical imaging remains challenging due to data privacy concerns, legal constraints, and the limited availability of diverse, high-quality datasets across healthcare institutions\cite{refK12}. Individual organizations often possess insufficient data, leading to model overfitting and reduced generalizability. Moreover, centralizing healthcare data may violate privacy regulations such as Health Insurance Portability and Accountability Act (HIPAA) and General Data Protection Regulation (GDPR)\cite{refK1}, restricting inter-institutional collaboration \cite{ref11}.
\par To address these challenges, researchers and medical institutions have explored Federated Learning (FL) \cite{ref11}, which allows collaborative model training without sharing raw data. Each participant updates a local model, which is aggregated by a central server. FL benefits medical imaging AI by enabling cross-institutional collaboration and preserving privacy, especially when sample sizes are limited. The aggregated global model often outperforms local models, though FL still faces issues with data heterogeneity and communication overhead \cite{ref17}.
\par Data heterogeneity\cite{refK21, refK22}occurs when datasets from different locations vary in distribution, quality, or quantity, hindering the global model's generalization. Additionally, large model updates can cause significant communication overhead in environments with many devices or slow networks. Strategies like model compression, tailored models, and efficient aggregation (FedAvg, adaptive federated optimization)\cite{refK17, refK18} have been applied to address these issues and improve FL performance. However, data heterogeneity and communication overhead\cite{refK19, refK20} remain bottlenecks. \cite{ref24}.\\
Existing studies \cite{refE, refH,refK13,refK14} on FL in medical imaging focus on inter-institutional privacy but often neglect data heterogeneity and communication overhead. This work proposes a unified solution by applying FL to non-IID medical data and using Top-K gradient sparsification to address these issues. Section 2 presents the problem formulation, Section 3 details the methodology based on the Flower framework and sparsification, Section 4 discusses datasets and implementation, and Section 5 reports the results. The paper concludes with key findings and future research directions.
\section{Problem Formulation}
\label{sec:problem}
Assume \( N \) hospitals, each with private data, collaborate in FL to train a shared global model without data sharing. The global model update in FL is given by:
\begin{equation}
    w^{t+1} = \sum_{i=1}^{N} p_i \, w^t - \eta \sum_{i=1}^{N} p_i \, \nabla F_i(w^t)
\end{equation}
Here, \( w^t \) denotes the model parameters at the \( t \)-th communication round, and \( w^{t+1} \) represents the updated global parameters after round \( t+1 \). \( p_i \) is the weight of client \( i \), typically proportional to its data share among all clients. \( N \) denotes the total number of clients, \( \eta \) is the learning rate, and \( \nabla F_i(w^t) \) is the gradient of the local loss function \( F_i(w) \) at client \( i \), evaluated at \( w^t \).

The first term, \( \sum_{i=1}^{N} p_i \, w^t \), computes the weighted average of the model parameters across all clients. In contrast, the second term, \( -\eta \sum_{i=1}^{N} p_i \, \nabla F_i(w^t) \), aggregates the gradient updates from each client, weighted by their data fraction, and scales the updates by the learning rate \( \eta \).

Due to the \textbf{heterogeneity of data} (i.e., clients having non-IID data), the model updates can vary significantly, leading to slower convergence and suboptimal performance. The global loss function is given by:
\begin{equation}
    \arg\min_{w} L(w) = \sum_{i=1}^{N} \left( \frac{|D_i|}{|D|} \right) \times \mathbb{E}_{x \sim D_i} \left[ l_i(w; x) \right]
\end{equation}

Where \( L(w) \) denotes the global loss function aggregating the local losses from all clients, \( N \) is the total number of clients, \( D_i \) is the dataset of client \( i \), and \( |D_i \) represents the number of data points in that dataset. \( |D| \) is the total data across all clients, and \( \frac{|D_i|}{|D|} \) indicates the weight of client \( i \)’s contribution to the global loss. \( F_i(w) = \mathbb{E}_{x \sim D_i} [ l_i(w; x) ] \) is the expected local loss for samples \( x \) from client \( i \)’s dataset, where \( l_i(w; x) \) is the loss for a given sample under model parameters \( w \).

This weighted sum ensures clients with larger datasets have greater influence on the global loss. Minimising this loss optimises model parameters \( w \) for overall performance. However, due to data heterogeneity, the communication cost can be high, as each client sends potentially large and varied updates. To address this, Top-k gradient sparsification transmits only a fraction \(K \in (0,1)\) of the largest updates per round, reducing the communication cost to \( \text{Comm\_cost}^{\text{sparse}} = N \times \text{Bytes}(w_{\text{sparse}}) \), meaning only \(k\%\) of updates are sent instead of the full model. Additionally, Top-k gradient sparsification helps mitigate the effects of data heterogeneity by prioritizing the most significant updates, leading to more stable convergence despite the variations in client data.
\section{Methodology}
\label{sec:Methodology}

Data heterogeneity in FL refers to variations in data distribution across clients, such as differences in labels, features, or sample sizes, which can affect convergence and generalization. To simulate non-IID medical imaging data, we partitioned datasets using a Dirichlet distribution, creating imbalanced splits that reflect typical heterogeneity in healthcare data. Each client's data partition \(P_i\) is drawn as \(P_i \sim \text{Dir}(\alpha)\), controlling data imbalance.
\begin{equation}
P(X_1, X_2, \dots, X_K; \alpha_1, \alpha_2, \dots, \alpha_K) = \frac{1}{B(\alpha)} \prod_{i=1}^{K} X_i^{\alpha_i - 1}
\end{equation}
To express this in a more detailed and comprehensive manner, let's first explicitly define all the terms involved:\\
$\prod_{i=1}^{K} X_i^{\alpha_i - 1}=X_1^{\alpha_1 - 1} \times X_2^{\alpha_2 - 1} \times \dots \times X_K^{\alpha_K - 1}$ is a product of terms, each of which is a fraction \( X_i \) raised to the power of \( \alpha_i - 1 \), each \( X_i \) represents the fraction of data assigned to the client \( i \),  
and \( B(\alpha) \) is the multivariate Beta function (the normalising constant), and it is defined as:
\begin{equation}
B(\alpha) = \prod_{i=1}^{K} \frac{\Gamma(\alpha_i)}{\Gamma\left(\sum_{i=1}^{K} \alpha_i\right)}
\end{equation}
Where \( \Gamma(\cdot) \) is the Gamma function and \( N \) is the number of clients. Each client \( i \) is assigned a fraction \( X_i \) of the data, with \( \sum_{i=1}^{N} X_i = 1 \). The concentration parameter \( \alpha = (\alpha_1, \alpha_2, \dots, \alpha_N) \) controls the data distribution, with each \( \alpha_i \) determining the data imbalance for client \( i \). 
\begin{figure} 
\centering
\includegraphics[width=0.57\textwidth, clip]{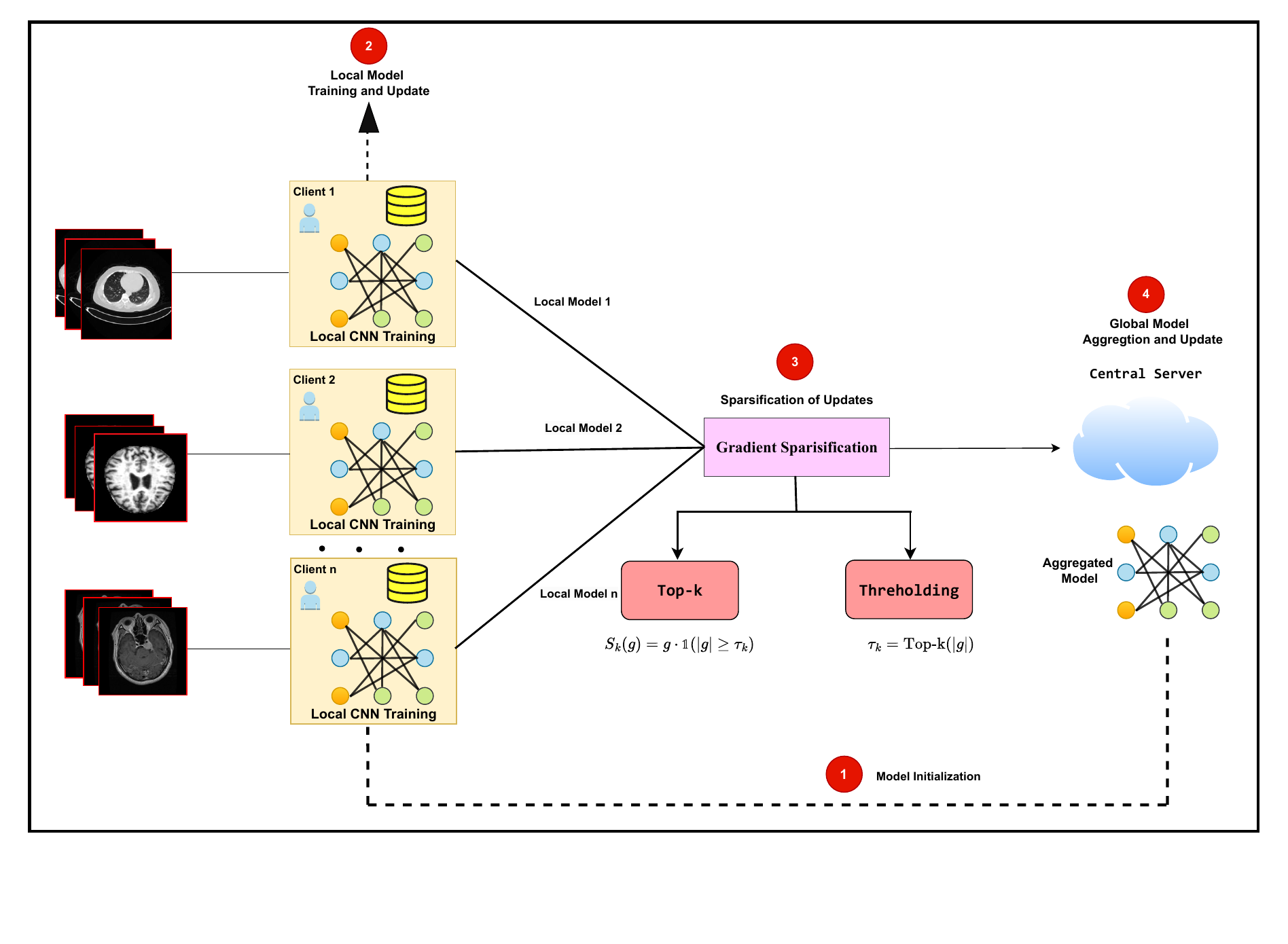}
\caption{\textbf{CFL-SparseMed} Architecture:.}
    \label{fig:brain_accuracy}
\end{figure}
The architecture in Figure \ref{fig:brain_accuracy} shows a system where clients train local CNN models with own local data. To minimise communication overhead, Top-k sparsification sends only the most significant gradients to a central server, which aggregates them to improve the global model. The updated model is then redistributed for further training, ensuring effective learning and data privacy.\\
Sending only Top-k gradient updates reduces communication costs and mitigates data heterogeneity by prioritizing key updates and minimizing outlier influence, enhancing convergence and generalization. The CFL-SparseMed algorithm is in \textbf{Algorithm 1}.
\begin{algorithm}[t!]
\caption{\textbf{CFL-SparseMed with ToP-K Sparsification}}
\label{alg:fed_topk}
\begin{algorithmic}[1]
\small
\STATE \textbf{Input:} Number of communication rounds \(T\), number of clients \(N\), number of local epochs \(E\), learning rate (\(\eta\)), sparsification rate (K), Dirichlet parameter (\(\alpha\)) 
\STATE \textbf{Server executes:}
\FOR{\(t=0,1,\ldots,T-1\)}
    \FOR{each client \(i\) (randomly selected in parallel)}
        \STATE Send global model \(w^t\) to client \(i\)
        \STATE \(w_i^t \gets\) \textbf{Client Local Training}(\(i, w^t\))
    \ENDFOR
    \STATE \textbf{Aggregate global model:}
    \STATE \( w^{t+1} = \sum_{i \in p_t} \left( \frac{|D_i|}{|D|} \right) w_i^t \)
\ENDFOR
\vspace{0.05cm}
\STATE \textbf{Client Local Training(\(i\)):}
\STATE \( w_i^t \gets w^t \)
\FOR{\(e=1,2,\dots,E\)}
    \FOR{each batch \(b=\{x,y\}\) from \(D_i\)}
        \STATE Compute loss: \( \ell_{\text{sup}} \gets \) Cross Entropy Loss(\( f(x),y \))
        \STATE Compute gradient: \( g \gets \nabla w_i^t \)
        \STATE \textbf{Top-k Sparsification:} Keep only the Top-k elements, denoted as \( g_{\text{top-k}} \)
    \ENDFOR
    \STATE \textbf{Update local model:} \( w_i^t \gets w_i^t - \eta g_{\text{top-k}} \)
\ENDFOR
\STATE Return \( w_i^t \) to server
\STATE \textbf{Output:} The final global model (\(w^t\))
\end{algorithmic}
\end{algorithm}
\section{Results and Discussions}
\label{sec:Results}
This study uses three openly accessible medical imaging datasets: Brain MRI \cite{ref33} (2,101 images, 3 classes), Alzheimer's \cite{ref34} (8,000 images, 4 classes), and Lung cancer \cite{ref35} (3,609 images), resized to 128$\times$128 pixels. Non-IID data is partitioned using a Dirichlet distribution (\(\alpha = 0.3, 0.6\)) across three clients (N=3).\\
FL framework Flower \cite{refE} with CNN500k trains via SGD (0.01 learning rate, Cross-Entropy Loss) for 5 epochs per client. Top-k sparsification with 
\(K = \{0.1, 0.2, 0.3, 0.4\}\) reduces communication overhead by sending only most significant gradients.\\
The global model trains for 200 rounds with full client participation, evaluated on client test sets on an NVIDIA A100 GPU.
\begin{table*}
\centering
\small
\resizebox{\textwidth}{!}{%
\begin{tabular}{|p{2.6cm}|p{1.7cm}|p{1.7cm}|p{1.4cm}|p{3.5cm}|p{3.5cm}|}
\hline
\textbf{Dataset} & \textbf{\#Clients} & \textbf{($\alpha$
)} & \textbf{K\%} & \textbf{Top-1 Acc. (\%) ($\alpha = 0.3$
)} & \textbf{Top-1 Acc.(\%)($\alpha = 0.6$
)} \\
\hline
\multirow{4}{*}{\textbf{BRAIN MRI\cite{ref33}}} & 3 clients & 0.3 / 0.6 & 0.1\% & 53.76 ± 0.5\% & 66.42 ± 0.5\% \\
                                    & 3 clients & 0.3 / 0.6 & 0.2\% & \textbf{\textcolor{green}{58.81 $\pm$ 0.5}}\% & \textbf{\textcolor{green}{68.88 $\pm$ 0.5}}\% \\
                                    & 3 clients & 0.3 / 0.6 & 0.3\% & 53.96 ± 0.5\% & 68.31 ± 0.5\% \\
                                    & 3 clients & 0.3 / 0.6 & 0.4\% & 51.34 ± 0.5\% & 67.81 $\pm$ 0.5\% \\
\hline
\multirow{4}{*}{\textbf{ADNI\cite{ref34}}}& 3 clients & 0.3 / 0.6 & 0.1\% & 77.93 ± 0.5\% & 82.14 $\pm$ 0.5\% \\
 
                                              & 3 clients & 0.3 / 0.6 & 0.2\% & \textbf{\textcolor{green}{80.20 $\pm$ 0.5}}\% & \textbf{\textcolor{green}{84.02 $\pm$ 0.5}}\% \\
                                              & 3 clients & 0.3 / 0.6 & 0.3\% & 79.69 $\pm$ 0.5\% & 80.46 ± 0.5\% \\
                                              & 3 clients & 0.3 / 0.6 & 0.4\% & 78.86 ± 0.5\% & 78.18 ± 0.5\% \\
\hline
\multirow{4}{*}{\textbf{LUNG CANCER\cite{ref35}}} & 3 clients & 0.3 / 0.6 & 0.1\% & \textbf{\textcolor{green}{96.13 $\pm$ 0.5}}\% & \textbf{\textcolor{green}{98.23 ± 0.5}}\% \\
                              & 3 clients & 0.3 / 0.6 & 0.2\% & 95.56 ± 0.5\% & 97.11 ± 0.5\% \\
                              & 3 clients & 0.3 / 0.6 & 0.3\% & 95.75 ± 0.5\% & 97.01 ± 0.5\% \\
                              & 3 clients & 0.3 / 0.6 & 0.4\% & 95.58 ± 0.5\% & 96.15 ± 0.5\% \\
\hline
\end{tabular}
}
\caption{CFL-SparseMed Evaluation: Accuracy vs. Sparsification on three datasets with with $\alpha$ = 0.3, 0.6 .}
\end{table*}
The \textbf{CFL-SparseMed} evaluation results presented in Table I show the impact of different top-k sparsification rates (\(K\)) and the corresponding top-1 test accuracy for each dataset.\\ 
 The Brain MRI dataset peaks at 58.81\% accuracy at \( K = 0.2 \), dropping to 51.34\% at \( K = 0.4 \). Alzheimer's reaches 80.20\% at \( K = 0.2 \), then drops to 78.86\% at \( K = 0.4 \). Accuracy for the Lung cancer dataset peaks at 96.13\% at \( K = 0.1 \), with \(\alpha = 0.3\) and remains stable for other sparsification rates.\\
 These results highlight the need for dataset-specific sparsification, with \textbf{CFL-SparseMed} balancing accuracy and communication efficiency. The stable accuracy of Lung cancer suggests robust features, while sparsification reduces performance in the more complex Brain MRI and Alzheimer's datasets by limiting subtle pattern capture. 
 \begin{figure*}[t!]
  \centering
  \includegraphics[scale=0.28]{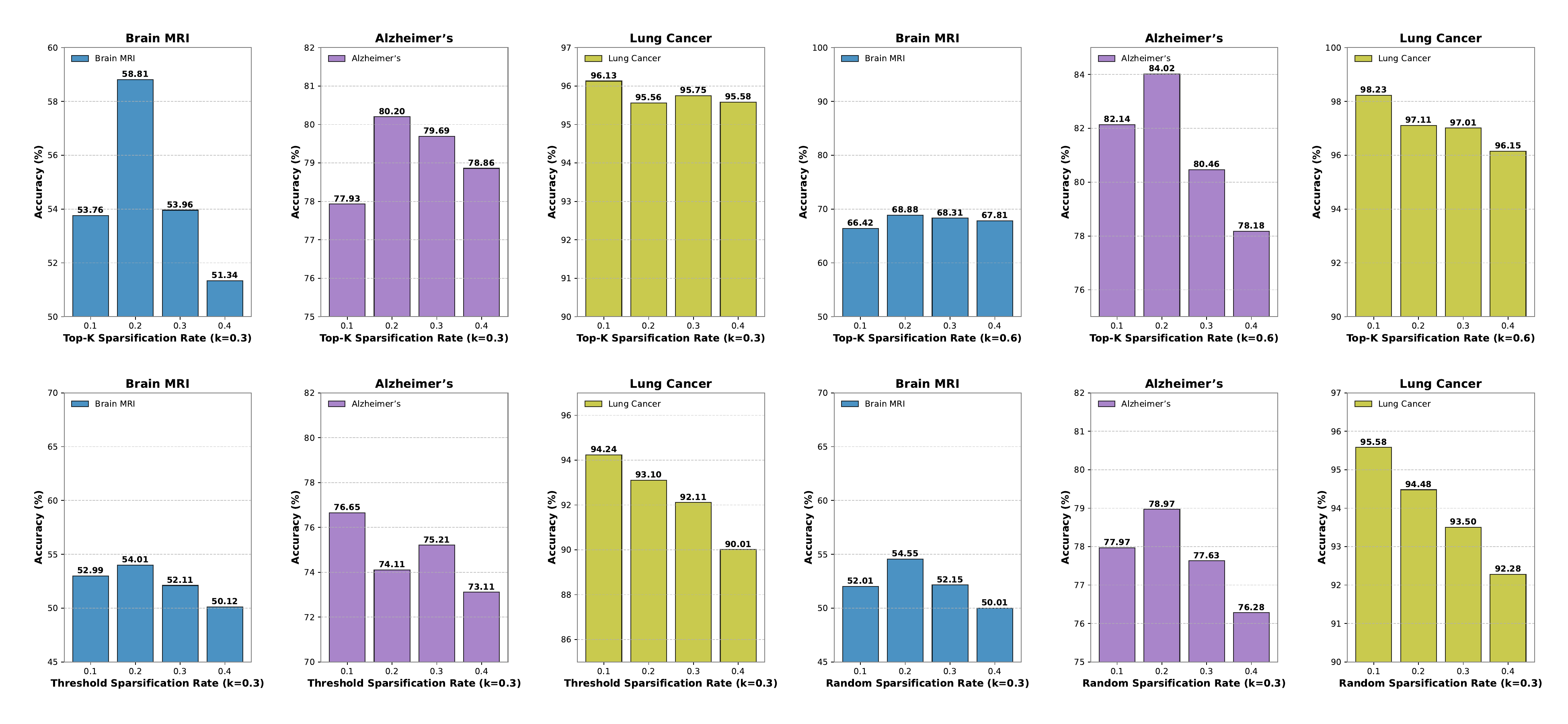} 
  \caption{Accuracy Vs Sparsity (200 rounds) for Proposed, Threshold, and Random methods with $\alpha$ = 0.3 on three datasets.}
  \label{Fa2}
\end{figure*}
Under the same experimental setup, we test our proposed method by increasing the concentration parameter (\(\alpha= 0.6 \)), leading to more balanced data partitioning and better model update aggregation. This enables the global model to learn more efficiently across clients.
Results in Table I show that \( \alpha = 0.6 \) outperforms \( \alpha = 0.3 \) across all sparsification rates (K = 0.1, 0.2, 0.3, 0.4), with the Lung dataset achieving the highest accuracy of 98.23\%, highlighting improved convergence and accuracy.\\
Subsequent trials were conducted using two existing sparsification techniques: threshold-based and random sparsification. For threshold-based sparsification, the sparsified update \( \tilde{s}_i \) is:
\[
\tilde{s}_i =
\begin{cases}
s_i & \text{if } |s_i| \geq \tau \\
0 & \text{if } |s_i| < \tau
\end{cases}
\]
Where \( \tau \) is the threshold (0.1, 0.2, 0.3, 0.4). Results presented Figure \ref{Fa2} across all the three datasets from Brain, Alzheimer and Lung show that lower thresholds (\( \tau = 0.1, 0.2 \)) improve accuracy, while higher thresholds degrade performance.
For random sparsification, varying rates (K = 0.1, 0.2, 0.3, 0.4) were tested. The sparsified update is:
\[
\tilde{s}_i =
\begin{cases}
s_i & \text{if } i \in \text{random subset} \\
0 & \text{otherwise}
\end{cases}
\]
Lower rates preserved more gradient information, improving performance, while higher rates led to declines due to information loss.
To further validate our method, we compared it against these techniques. The results presented in Figure\ref{Fa2} shows that our approach outperforms consistently against both threshold-based and random sparsification across all sparsification rates and with a higher heterogeneity rate of  \( \alpha = 0.3 \).\\
Compared to baseline methods in Table 2, \textbf{CFL-SparseMed} outperforms FedAvg (54.55\%), Moon (57.26\%), and FedProx (56.19\%) in accuracy, achieving 58.81\% with 200 communication rounds(CR). It also converges faster with convergence time(CT) (9000 secs) than Moon (36000 secs) and FedProx (43200 secs), offering a strong balance between performance and communication efficiency.
\begin{table}[H]
\centering
\small
\renewcommand{\arraystretch}{1.4} 
\begin{tabular}{|p{1.33cm}|p{1.9cm}|p{1.7cm}|p{0.4cm}|p{1.78cm}|}
\hline
\textbf{Method}  & \textbf{\#CR} & \textbf{CT(secs)} & \textbf{$\alpha$} & \textbf{Top-1-Aucc.} \\ \hline
\textbf{FedAvg\cite{ref17}}    & 1000                     & 14,400                     & 0.3                     & 54.55 $\pm$ 0.2\%               \\ \hline
\textbf{MOON\cite{ref17}}       & 1000                    & 36,000                      & 0.3                     & 55.20 $\pm$ 0.2\% \\ \hline
\textbf{FedProx\cite{ref17}}    & 1000                   & 43,200                      & 0.3                     & 56.19 $\pm$ 0.2\%               \\ \hline
\textbf{OURS} & \textcolor{green}{\textbf{200}}                   & \textcolor{green}{\textbf{9000}}                      & 0.3                     & \textcolor{green}{\textbf{58.81} $\pm$ 0.2}\%               \\ \hline
\end{tabular}
\caption{Comparison of \textbf{CFL-SparseMed} with Baselines.}
\end{table}
\section{Conclusion}
\label{sec:Conclusion}
We propose \textbf{CFL-SparseMed}, a communication-efficient FL approach for medical imaging. It uses Top-k gradient selection to reduce communication overhead while maintaining high model performance across diverse datasets. By transmitting only significant updates, \textbf{CFL-SparseMed} ensures efficient communication in bandwidth-limited environments. Compared to baselines, \textbf{CFL-SparseMed }converges faster, making it ideal for real-world federated medical settings with limited resources. It is well-suited for distributed medical data analysis in hospitals and clinics.\\
\textbf{CFL-SparseMed} approach addresses heterogeneity, but challenges remain in scalability, as Dirichlet simulations may not represent real medical data. Future work could integrate contrastive loss and adaptive techniques to improve efficiency, generalization, and scalability in large-scale FL for medical imaging.

\textbf{Acknowledgments}
No funding was received for this study. The authors have no relevant interests to disclose.
\bibliographystyle{IEEEbib}
\bibliography{strings,refs}

\end{document}